\documentclass[openacc]{rstransa}
\usepackage{cuted,adjustbox}
\usepackage{multirow,multicol}
\usepackage{booktabs,colortbl}
\usepackage{xcolor}
\usepackage[labelformat=simple]{subcaption}

\usepackage{tikz}
\usepackage{pgfplots} 
\usepackage{circledsteps}
\usepackage{listings}
\usepackage{xspace}
\usepackage{tcolorbox}
    
\tcbuselibrary{listings}
\newcommand\myCircled[2][]{\ifmmode
\Circled[fill color=black,inner color=white,#1]{\mathsf{#2}}
\else
\Circled[fill color=black,inner color=white,#1]{\sffamily#2}
\fi
}
\usepackage{orcidlink}
\usepackage{url}
\definecolor{rwth_blue1}{RGB}{87,171,39}
\definecolor{rwth_blue2}{RGB}{141,192,96}
\definecolor{rwth_blue3}{RGB}{184,214,152}
\definecolor{rwth_blue4}{RGB}{221,235,206}
\definecolor{rwth_blue5}{RGB}{242,247,236}

\newcommand{\pieslice}[6][black!10]{
  \pgfmathparse{#3/#2*360}
  \let\a\pgfmathresult
  \pgfmathparse{#4/#2*360}
  \let\b\pgfmathresult

  \pgfmathparse{0.5*\a+0.5*\b}
  \let\midangle\pgfmathresult

  \draw[fill=#1] (0,0) -- (\a:1) arc (\a:\b:1) -- cycle;

  \node[label=\midangle:{\small#6}] at (\midangle:1) {};

  \pgfmathparse{min((\b-\a-10)/110*(-0.3),0)}
  \let\temp\pgfmathresult
  \pgfmathparse{max(\temp,-0.5) + 0.8}
  \let\innerpos\pgfmathresult
  \pgfmathparse{(\b-\a)/3.6} 
  \let\percentage\pgfmathresult
  \node at (\midangle:\innerpos) {\small\pgfmathprintnumber[fixed,precision=1]{\percentage}\%};
}

\newcommand{\pie}[2][{{"black!10"}}]{
  \pgfmathparse{dim(#1)} 
  \let\paletteDim\pgfmathresult
  \newcounter{colourIndex}

  \newcounter{total}
  \foreach \val/\name in #2 {
    \addtocounter{total}{\val}
  }

  \newcounter{a}
  \newcounter{b}
  \foreach \val/\name in #2 {
    \setcounter{a}{\value{b}}
    \addtocounter{b}{\val}

    \pgfmathparse{#1[\thecolourIndex]}
    \let\colour\pgfmathresult

    \pieslice[\colour]{\thetotal}{\thea}{\theb}{\val}{\name}

    \stepcounter{colourIndex}
    \ifnum \thecolourIndex=\paletteDim \setcounter{colourIndex}{0}\fi
  }
}

\usepackage{background}
\usepackage[usestackEOL]{stackengine}
\setstackgap{L}{\normalbaselineskip}
\SetBgContents{\color{blue}{\tiny \Longstack{PREPRINT - accepted for the Philosophical Transactions of the Royal Society A}}}
\SetBgPosition{3.5cm,2cm}
\SetBgOpacity{1.0}
\SetBgAngle{0}
\SetBgScale{1.8}
\titlehead{Research}

\begin{document}

\title{Exploiting the Lock: Leveraging MiG-V's Logic Locking for Secret-Data Extraction}

\author{
Lennart M. Reimann$^{1}$\orcidlink{0009-0003-5825-2665}, Yadu Madhukumar Variyar$^{1}$, Lennet Huelser$^{1}$, Chiara Ghinami$^{1}$\orcidlink{0009-0009-1430-2329}, Dominik Germek$^{2}$\orcidlink{0000-0003-3812-727X} and Rainer Leupers$^{1}$\orcidlink{0000-0002-6735-3033}}

\address{$^{1}$RWTH Aachen University, Germany, \{lennart.reimann, madhukumarvariyar, huelser, ghinami, leupers\}@ice.rwth-aachen.de \\
$^{2}$Corporate Research, Robert Bosch GmbH, Germany, dominik.germek@de.bosch.com}

\subject{xxxxx, xxxxx, xxxx}


\corres{Lennart M. Reimann\\
\email{lennart.reimann@ice.rwth-aachen.de}}

\begin{abstract} 
The MiG-V was designed for high-security applications and is the first commercially available logic-locked RISC-V processor on the market. In this context logic locking was used to protect the RISC-V processor design during the untrusted manufacturing process by using key-driven logic gates to obfuscate the original design. Although this method defends against malicious modifications, such as hardware Trojans, logic locking's impact on the RISC-V processor's data confidentiality during runtime has not been thoroughly examined.
In this study, we evaluate the impact of logic locking on data confidentiality. By altering the logic locking key of the MiG-V while running SSL cryptographic algorithms, we identify data leakages resulting from the exploitation of the logic locking hardware. We show that changing a single bit of the logic locking key can expose 100\% of the cryptographic encryption key. This research reveals a critical security flaw in logic locking, highlighting the need for comprehensive security assessments beyond logic locking key-recovery attacks.

\end{abstract}


\begin{fmtext}
\section{Introduction}
The modern Integrated Circuit (IC) design and fabrication process is driven by the need for rapid time-to-market\end{fmtext}
\maketitle
\noindent and reduced design costs. This has resulted in a horizontal business model where IC design houses depend on third-party Intellectual Property (IP) and outsource fabrication to off-site foundries. The inclusion of untrusted parties in both the design and fabrication stages has introduced numerous security concerns, including IP piracy and the insertion of malicious modifications, commonly referred to as hardware Trojans. A promising solution to protect the IC integrity is Logic Locking (LL), which involves the insertion of additional key-controlled logic to a gate-level netlist, making the IP design dependent on a secret key~\cite{mentor, disquisition}. Hereby, the key is known only to the original IP owner. This approach keeps the design concealed from untrusted external designers and foundries. The security of LL hinges on the assumption that a malicious entity must first uncover the activation key before being able to reverse engineer the design and implement a meaningful (design-dependent) hardware Trojan.
\begin{figure}
    \centering
    \includegraphics[angle=-90, width=0.7\columnwidth]{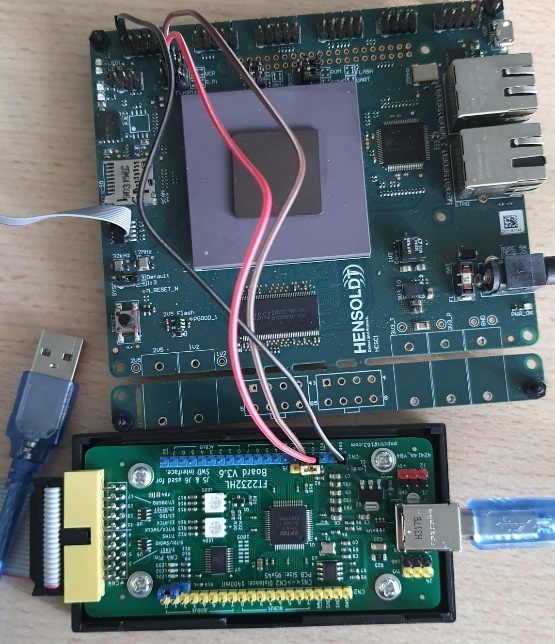}
    \caption{The MiG-V evaluation board with a connected JTAG adapter.}
    \label{fig:mig-v-jtag}
    \vspace{-0.4cm}
\end{figure}
Hensoldt Cyber (HC) GmbH designed the MiG-V, a RISC-V processor developed in Germany targeting high-security applications~\cite{mig-v, mig-v2, risc-v-news} (see Figure~\ref{fig:mig-v-jtag}). In the MiG-V, HC combined a secure hardware design with a formally verified software kernel, seL4, to secure the entire hardware-software stack. In this context, LL is used to address threats such as malicious modifications during the entire process in the external design house and foundry. As the LL key is concealed from the untrusted parties, the IP's behavior cannot be easily derived from the hardware description, preventing the incorporation of malicious modifications, such as Denial-of-Serice (DoS) hardware Trojans, within obscured segments of the hardware's functionality~\cite{dmux2022}. 

While current LL policies aim to protect ICs during fabrication, they often fail to address their impact on a crucial security aspect: confidentiality. As LL is applied after the chip is fully designed, enforced security properties might be broken again by the newly introduced logic. Incorporating key-controlled logic within the circuit can create substantial vulnerabilities. This crucial oversight has not been thoroughly investigated in existing studies, resulting in important security risks remaining unaddressed.

To address this research gap, our study focuses on evaluating how LL affects the confidentiality of the MiG-V processor during its in-field operation. In this context, we evaluate its effects on the program flow of the open-source OpenSSL cryptographic software~\cite{openssl}, and how sensitive data, such as the cryptographic encryption keys can be leaked to an adversary. \textit{The encryption key is used for data encryption and is distinct from the logic-locking key.} Numerous studies have already focused on attacks that threaten the security of LL, such as Boolean Satisfiability (SAT)-based attacks and fault injection vulnerabilities in processors\cite{sat_attack1, sat_attack2}. However, our objective is \textit{not to break the logic-locking scheme}, \textit{but to pinpoint direct data leakages specifically caused by LL} within the MiG-V.

To assess the impact of the MiG-V's Inter-Lock logic-locking scheme on the confidentiality of cryptographic software running on the processor, we examined whether alterations to the LL key could cause the MiG-V to output the sensitive encryption key instead of the encrypted ciphertext. Additionally, we designed hardware Trojans that enable an adversary to change the LL key in a manufactured processor. The key contributions of this work are as follows:
\begin{enumerate}
    \item The evaluation of the impact of LL in a commercially available RISC-V core on the confidentiality of cryptographic software.
    \item The development and implementation of a hardware Trojan to allow changing the LL key after activation. 
    \item An exemplary exploitation of LL to leak data.
\end{enumerate}
The rest of the paper is organized as follows: Section 2 introduces the Inter-Lock LL technique, the attack model, and the MiG-V hardware system under analysis. Section 3 reviews related work. Section 4 details the methodology, and Section 5 presents the experimental evaluation and results. Section 6 discusses the findings, and Section 7 concludes the paper.

\section{Background}
The following introduces logic locking, the MiG-V processor, and the attack model. 

\begin{figure}[t]
    \centering
    \includegraphics[origin=c, width=0.95\textwidth]{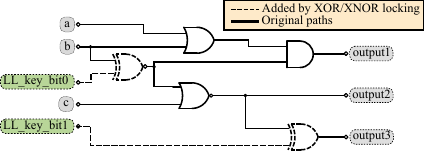}
    \caption{Placing XOR/XNOR logic locking key gates using an XOR/XNOR logic locking scheme.\label{fig:epic}}
    \vspace{-0.4cm}
\end{figure}

\paragraph{\textbf{Logic Locking}}

Logic locking alters hardware designs by inducing functional changes that are dependent on a secret activation key. This mechanism ensures that only a correct key preserves the original functionality, while an incorrect key results in erroneous outputs~\cite{evolution, provably_lolo}. Since LL introduces functional and structural changes in the design, it not only prevents the utilization of the correct functional behavior in case of a missing correct key but also induces a more fundamental impact: the complexity of reverse-engineering a locked design \textit{increases}.

A significant branch of LL is dedicated to the application of XOR/XNOR gates for locking ~\cite{Sisejkovic2023AttacksAndSchemes}. An example of this mechanism is presented in Fig.~\ref{fig:epic}. Here, the inserted XOR and XNOR gates are controlled via the key inputs $lolo\_key\_bit0$ and $lolo\_key\_bit1$. For $lolo\_key\_bit1=0$, the second input of the XOR gate is buffered to its output, thus preserving the original design's functionality. If $lolo\_key\_bit1=1$, the second input is inverted, resulting in erroneous behavior. Similarly, the XNOR key gate preservers the input value for $lolo\_key\_bit1=1$. To avoid a one-to-one mapping between the correct key value and the type of the corresponding gate, the locking mechanism can insert inverters at the output of the XOR/XNOR gates. Thus, in a structural sense, the attacker must guess if the inverter is part of the original design or the locking mechanism.  

This fundamental mechanism has been integrated into various XOR/XNOR-based LL schemes~\cite{10.1145/2228360.2228377,7362173,6616532,9214869}. A differentiating factor is provided by the specifics of the insertion strategy of the key-controlled gates that serve certain security objectives, such as introducing changes at random locations, maximizing the functional corruptibility, creating a high interdependence between the gates, and others~\cite{Sisejkovic2023AttacksAndSchemes}. As a random insertion represents a superset of all strategies, further evaluations in this work are based on the EPIC scheme~\cite{epic}. EPIC inserts XOR/XNOR gates at random locations in the netlist.

\paragraph{\textbf{Reverse Engineering and Hardware Trojans}} The process of reverse engineering a design can be roughly divided into two steps\footnote{Assumption: reverse engineering starts from a gate-level netlist.}~\cite{8494896, Sisejkovic2023WorkingPrinciples, 10.1145/3287624.3288740}. First, the target hardware is partitioned into its structurally independent and functionally cohesive components. Second, the identification of the \textit{functionality} of the subcomponents and, afterward, the entire hardware is performed. As reverse engineering plays a critical role in understanding the target hardware and designing and injecting a design-dependent hardware Trojan, the objective of LL must be twofold: increasing the complexity of both the partitioning \textit{and} the functional identification. A method to address both objectives with LL was introduced in the scheme Inter-Lock, as discussed in the following.

\paragraph{\textbf{Inter-Lock}} In standard LL schemes, every component of a hardware design is individually considered for locking. This leads to an attack strategy in which each component is subsequently attacked. As a remedy, the Inter-Lock scheme was proposed~\cite{8791528,9088001}. Inter-Lock is a cross-module scheme-independent locking procedure that induces additional security dependencies between selected locked components in a design, yielding two outcomes. First, the locked and interconnected components become functionally dependent and structurally cohesive. Second, the functional activation of each component relies on the activation of codependent components. The former outcome increases the complexity of performing partitioning. The latter forces an attacker to consider multiple components simultaneously to infer a correct key. Therefore, Inter-Lock caters to all major requirements of making reverse engineering harder to perform.

\paragraph{\textbf{The MiG-V Core}} The Inter-Lock scheme was brought to life in form of the MiG-V---the first logic-locked commercial processor~\cite{mig-v}. The core of the MiG-V is the 64-bit 6-stage, in-order, single-issue Ariane core~\cite{ariane}. The core implements the open-source RISC-V instruction set architecture. The MiG-V is protected through the application of Inter-Lock with a 1024-bit key using the random XOR/XNOR-based LL scheme. Hereby, only selected critical components of the core are locked. The reasoning is that a meaningful hardware Trojan requires a certain intelligence for activation. Thus, locking a carefully selected set of critical components serves as a protection layer against severe hardware Trojans.

The MiG-V SoC was not developed by us, but by the Hensoldt Cyber GmbH. We did not decide on the usage of the logic locking scheme nor the length of the logic locking key for the manufactured board. However, the influence of logic locking on the processor's area and performance values is evaluated in ~\cite{more_migv_eval}.

\subsection{MiG-V Evaluation Board (Logic Locked RISC-V)}
\begin{figure}
    \centering
    \includegraphics[width=\columnwidth]{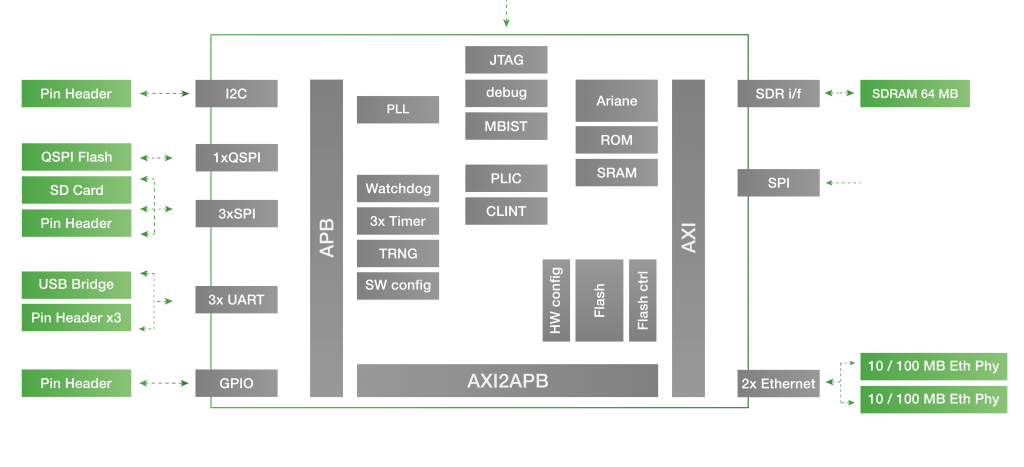}
    \caption{Overview of the MiG-V~\cite{mig-v-blockdiagram}.}
    \label{fig:mig-v-blockdiagram}
    \vspace{-0.4cm}
\end{figure}

The MiG-V implements a logic-locked 64-bit RISC-V processor using the Integer, Multiply, Atomic, and Compressed extensions. HC designed the MiG-V by modifying and improving the open-source processor Ariane~\cite{ariane}. The processor implements a Read-Only Memory (ROM), a static random access memory (SRAM), a synchronous dynamic random access memory (SDRAM), and an Instruction and Data Cache, as depicted in Figure~\ref{fig:mig-v-blockdiagram}. Further details about the memories important for this research are described in Table~\ref{tab:memory}.
\begin{table}[]
    \centering
    \begin{tabular}{c|c|c}
        Memory type & Size & Purpose \\ \hline
        ROM & 512kB  & Zero-stage-bootloader, operating systems\\ \hline
        Flash & 2MB & Bootloader, logic-locking key\\ \hline 
        SRAM & 1MB & Additional bootloaders, data, programs, operating systems \\ \hline
        SDRAM & up to 64MB & Additional bootloaders, data, programs, operating systems\\ \hline
    \end{tabular}
    \caption{Memory implementations in the MiG-V.}
    \label{tab:memory}
    \vspace{-0.4cm}
\end{table}
The LL key can be written into the flash memory using the JTAG MBIST interface, which has full access to the internal flash ports. The zero-stage bootloader loads the LL key from the flash memory to internal registers when restarting the evaluation board. Therefore, changes to the LL key within the evaluation board are only valid before restarting it. Due to this limitation, an attacker needs an additional Trojan, enabling runtime changes to the LL key, as discussed in subsequent sections.

The user can load programs into the SRAM and SDRAM via the UART or JTAG interface. Additionally, the JTAG interface offers the usage of an on-chip debugger, which allows loading the program and executing it cycle by cycle, while observing internal register and memory values. The UART interface can be used for simple I/O communication, such as outputting the content of C printf functions to a host computer. 

The MiG-V is logic-locked using the Inter-Lock mechanism with a 1024-bit key. The locking process was focused on the security-critical modules of the processor aiming to reduce the area and performance overhead caused by LL, while securing the hardware. The following modules were logic-locked.
\begin{enumerate}
\begin{multicols}{2}
    \item branch prediction
    \item instruction scan
    \item pc\_select
    \item instruction realign logic
    \item decoder
    \item ALU
    \item CSR buffer
    \item branch unit
    \item commit stage
    \item controller
    \item compressed decoder
    \end{multicols}
\end{enumerate}

Setting "incorrect" LL key bits results in flipped internal control or data bits due to the implemented XOR/XNOR key gates. The main focus of this work lays on the encrypted decoder and compressed decoder, as both modules are responsible for decoding the loaded instructions and setting the control signals in the processor. Even minimal bit flips resulting from changes to the LL key can cause misinterpretation of instructions and incorrectly set control signals for unintended instructions, thereby significantly altering the program flow. Each of the two decoder modules is logic-locked using 128 key gates.

\subsection{Attack Model}

The attackers' approach can be divided into two stages: 1. Analysis \& Trojan implementation, 2. Attack. First, they analyze the hardware description to determine malicious LL keys and implement a hardware Trojan near the LL key storage. Subsequently, they utilize the collected information to launch an attack on the manufactured and activated circuit.

\paragraph{\textbf{Analysis}}
In the analysis phase, we assume the adversary has access to the logic-locked netlist. This netlist can be obtained either by direct access to the design by working for the external design house or the foundry. Additionally, the adversary can be a rogue employee of the IP developer.  We also assume the adversary:
\begin{itemize}
    \item ... knows the location of the LL key storage.
    \item ... has minor knowledge about the design and the software that will run on it.
    \item ... can observe the outputs of the design either remotely or through direct chip access.
\end{itemize}

The adversary can manipulate the LL key inputs by tampering with the storage holding the LL activation keys or modifying the value before it reaches the key gates. The components near the storage of the LL key cannot be obfuscated by LL as the hardware to write the key needs to be functional even for a non-activated chip to allow for activating it\cite{taal}. Hardware Trojans, such as TAAL~\cite{taal}, can exploit vulnerabilities near the key storage to leak the key after activation. In this work, we developed a hardware Trojan capable of switching the original LL key with the malicious, so-called "Trojan keys".

\paragraph{\textbf{Attack}}
In the second phase, the adversary, now acting as an end-user, has access to an activated manufactured chip. By activating the designed hardware Trojan, the attacker can temporarily alter the key combination in the activated IC to gain access to sensitive data, such as encryption keys or user data. This research demonstrates that such access to sensitive information is enabled by LL. After extracting the desired data, the attacker can restore the original functionality by reapplying the correct LL key. The design of the hardware Trojan is further discussed in Section~\ref{ch:methodology}. 

\section{Related Work}
To the best of our knowledge, this work is the first to evaluate the influence of LL on the confidentiality property in secure hardware designs. Recent studies have demonstrated the exploitation of LL to compromise the integrity of neural accelerators during runtime. In such cases, LL is employed as a backdoor to diminish the accuracy of correct detections~\cite{lolo_ai_attack}.

The path sensitization method used in this research has been applied in other LL contexts as well. For instance, path sensitization has been utilized to assess the secure storage of LL key bits~\cite{10.1145/2228360.2228377}. This involves evaluating whether the LL key bits can be transmitted to the design's output using a calculated input sequence. Additionally, SAT-attacks---a prevalent set of key-retrieval attacks---solve Boolean satisfiability problems to determine how inputs propagate from primary inputs to primary outputs~\cite{sat_attack1, sat_attack2, sat_attack3, sat_on_lolo}. By retrieving the LL key bits, these attacks can unlock the netlist, enabling IP piracy, overproduction, and the insertion of hardware Trojans. If the same locked design is reused, the LL key can be employed to recover the design in subsequent manufacturing batches.

Moreover, machine learning techniques have been used to recover the original functionality of a locked design~\cite{lolo_ml,titan, omla}. Existing studies have primarily concentrated on whether the original design or the LL key can be retrieved.

However, the impact of LL itself on the security properties of the unlocked design has not been thoroughly analyzed. This research addresses this gap by evaluating \textit{whether LL can be exploited to leak sensitive data within a design.}

\section{Methodology}
\label{ch:methodology}

To assess the impact of LL on the MiG-V processor, we implemented a structured methodology that involves systematically altering the LL key and evaluating the resulting outputs from cryptographic applications.

\subsection{Evaluation Board Setup}
The MiG-V evaluation board is connected to a host computer using an FT2232HL adapter, as shown in Figure~\ref{fig:mig-v-jtag}. The adapter enables a JTAG and UART connection, which allows communication between the MiG-V and the host computer. 
First, we prepared the MiG-V processor to run a series of cryptographic applications in a controlled environment. 
We assume that, if logic locking endangers the security of highly secured algorithms, other software can be threatened as well.
We used the following OpenSSL~\cite{openssl} C programs for the evaluation: the encryption schemes AES-GCM-256, ChaCha and Camellia, and the hashing algorithm SHA-256. For the encryption schemes we labeled the encryption key as sensitive data, for the hashing scheme the plaintext is labeled. The applications' output of the MiG-V with the correct LL key serves as the baseline for our comparisons.
We developed a script to automate the process of changing individual bits of the 1024-bit LL key. This script systematically flips each bit, one at a time, ensuring that each unique key variation is tested independently. Before booting the processor for each test, the script automatically flipped one specific bit of the LL key while leaving the rest unchanged. This process was repeated for all 1024 bits.
After modifying a single bit of the LL key, the script loaded the cryptographic applications onto the processor, executing them sequentially. During the execution, the output from each application's printf operations, which write out the ciphertext via the UART output, was monitored and captured.

The UART output was then automatically compared to the baseline's output. Any tests that have an output deviating from the correct response are manually reviewed. Whenever a change in the output was observed, further manual inspection determined whether the altered ciphertext contained parts of the sensitive data. Each observed change was documented, noting which bit flip caused the deviation and the nature of the observed output change.
This methodology allowed for a comprehensive analysis of the influence of individual LL key bit flips on the MiG-V processor's confidentiality. By systematically testing each bit and manually inspecting the outputs, we aimed to identify potential security weaknesses introduced by the LL scheme.
\subsection{FPGA Setup}
As previously noted, the MiG-V evaluation board permits changes to the LL key only before the reset, preventing any attacks from occurring post-reset. The evaluation board provides simple read-and-write access to the LL key storage to ensure users have full access to all system components. However, in high-security applications, this access would be much more restricted, eliminating the possibility of reading from or writing to the LL key storage. 

However, to illustrate that even a MiG-V with restricted access to the logic locking key storage can be compromised, this work develops a hardware Trojan capable of changing the key during runtime, overcoming these restrictions. This Trojan can be implemented by an adversary in the design house or foundry, or by a rogue IP designer before a new version of the LL chip is manufactured. Furthermore, although LL aims to prevent Trojans, the surrounding hardware's security remains a concern, as it may not be logic-locked to enable circuit activation. The hardware Trojan represents a proof-of-concept for any attacks that require an adversary to change the logic locking key, although the IP designer restricted that access.

We implemented the complete logic-locked MiG-V and its peripherals on the AMD ZCU102 Evaluation Kit. To integrate and control a hardware Trojan within the MiG-V architecture, a new Control and Status Register (CSR), named mtrojanreg, was introduced. This register operates at the highest privilege level, providing seamless control over the Trojan. The CSR connects to the select input of the Trojan module, enabling an attacker to switch between the correct locking key and various attack keys effortlessly.

\begin{figure}
    \centering
    \includegraphics[width=\textwidth]{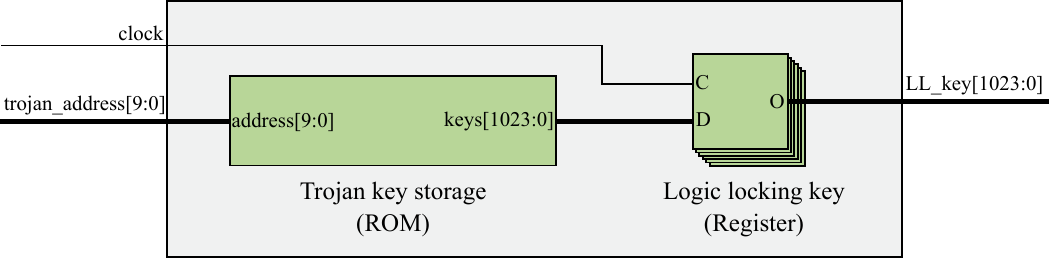}
    \caption{Implemented Trojan on the ROM. 1024 different adverse LL keys can be stored, addressable using 10-bit (trojan\_address[9:0]).}
    \label{fig:trojan}
    \vspace{-0.4cm}
\end{figure}
The Trojan utilizes a memory-based approach, leveraging a ROM to store the Trojan keys (see Figure~\ref{fig:trojan}). The design accesses the ROM using input addresses. The Trojan is designed to be triggered using internal signals within the SoC. Implemented in separate SystemVerilog files, the Trojan is imported into the design and instantiated at the top level of the SoC. For the purpose of synthesis and implementation, the Trojans were initially loaded with a set of 1024 attack keys. This number can be reduced after identifying the most effective keys responsible for the desired adverse behavior.
By incorporating the hardware Trojan, we can prove that an adversary can use Trojans to modify LL keys after manufacturing using software highlighting the vulnerability caused by LL. The software can be used to change the contents of the implemented CSR, thereby controlling the Trojan and starting the attack.

\section{Evaluation}
\label{ch:evaluation}

\subsection{Evaluation Board Results}
\begin{figure}
\begin{subfigure}{0.5\columnwidth}
    \centering
    \includegraphics[{width=0.9\textwidth}]{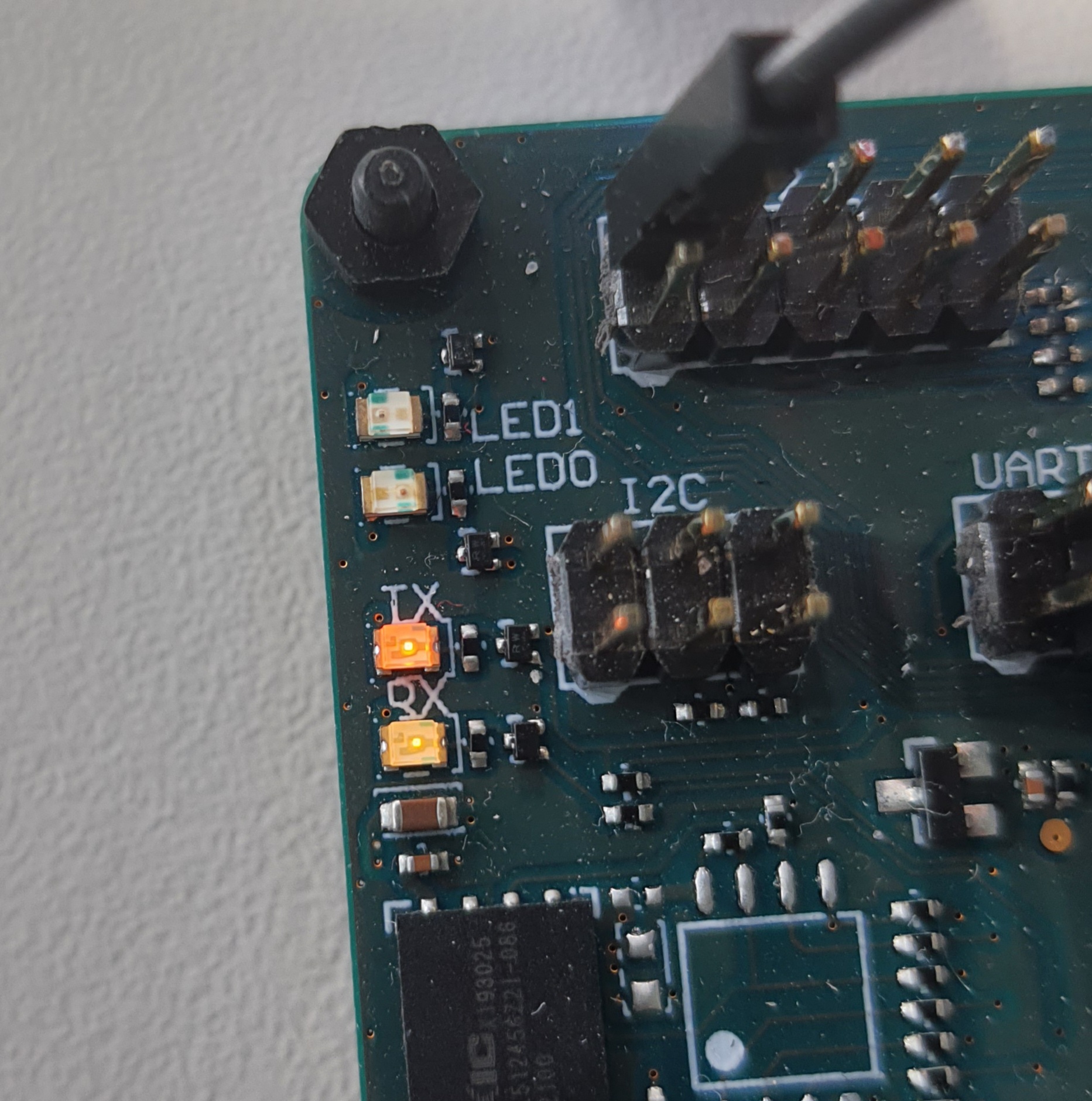}
    \caption{Host to MiG-V connection denied.}
    \label{fig:led_off}
\end{subfigure}
\begin{subfigure}{0.5\columnwidth}
    \centering
    \includegraphics[width=0.945\columnwidth]{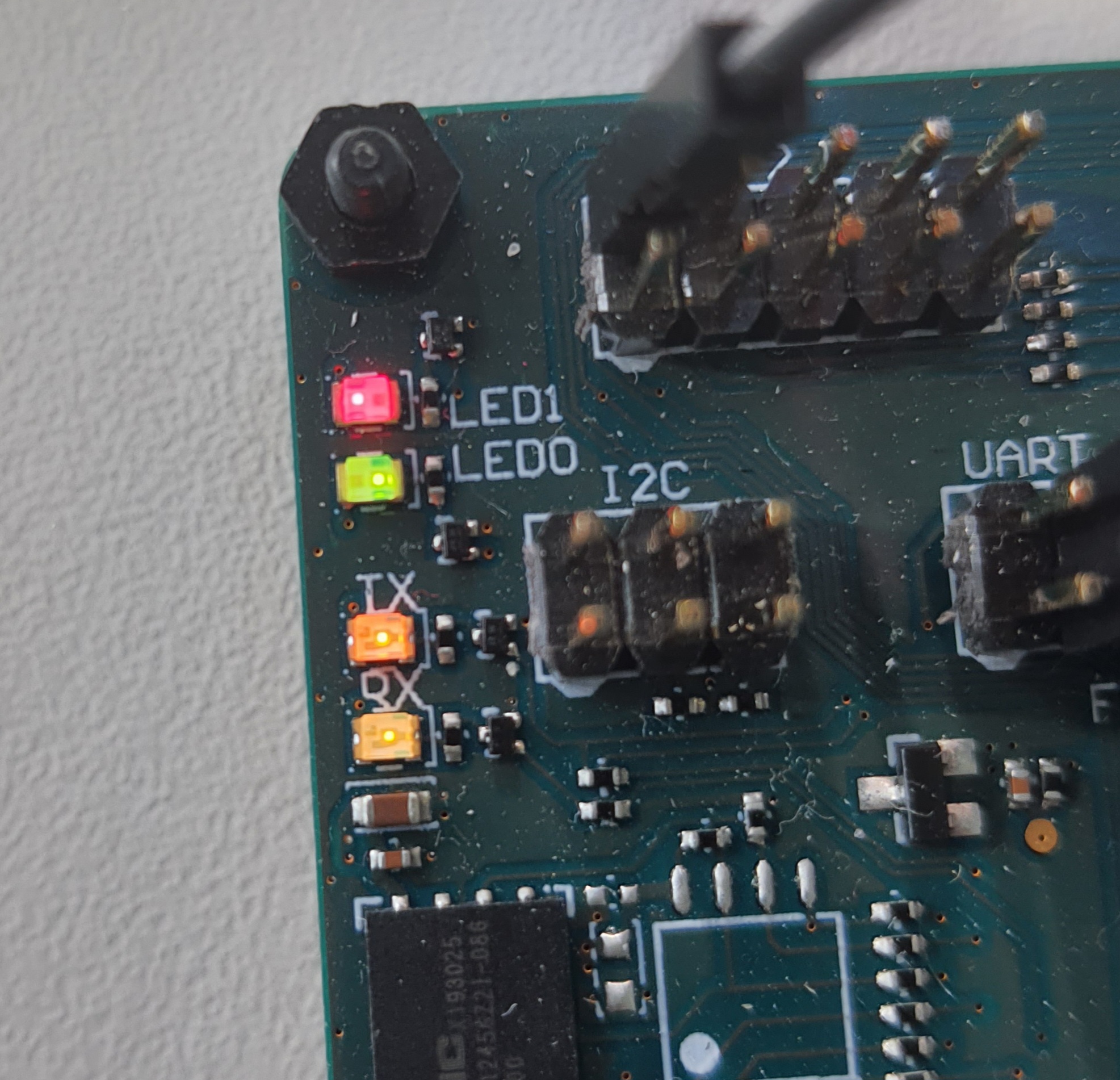}
    \caption{Host to MiG-V connection possible.}
    \label{fig:led_on}
\end{subfigure}
\caption{The MiG-V's board status LEDs signaling a valid Open-OCD connection.}
\label{fig:leds}
\vspace{-0.4cm}
\end{figure}

During the evaluation, we test 1024 bit manipulations, changing each bit of the 1024 LL key bit once. As a change in the LL key results in a change in the processor's functionality, many of the bit flips result in the processor not booting up any longer. A substantial portion of the bit manipulations result in a complete shutdown of the board's connection, and without the ability to use OpenOCD to load applications on the board, analyzing the impact of bit flips on information flow becomes infeasible. During the analysis, the LL key bit flips that resulted in an invalid connection can be identified by observing the status LEDs, as depicted in Figure~\ref{fig:leds}. 

An overview of the resulting changes caused by the bit flips is illustrated in Figure~\ref{fig:pie_chart}. 44\% of the bit flips result in an invalid connection. Approximately 64\% of the LL key bits caused a communication denial with the MiG-V. In our investigation of a potential Trojan key, we examined bit flips that led to changes in output. Out of 87 instances where altered outputs were detected, we found that in 27 cases, none of the four applications produced any output. These occurrences were categorized as changed output instead of no Picocom output, as they could still print text, but not variables. This included the absence of printed matrices, which store ciphertexts and hashes.

\def\palette{{"rwth_blue1","rwth_blue2","rwth_blue3","rwth_blue4",
    "rwth_blue5","brown!50!black!50","purple!50","lime!50!black!30"}}

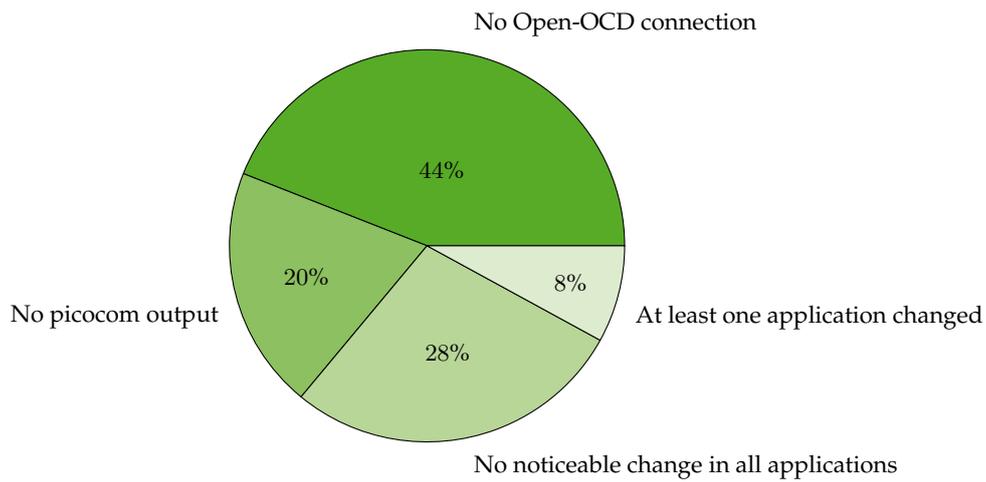
\begin{figure}
    \centering
    \begin{tikzpicture}[scale=2.6]
  \pie[\palette]{{44/No Open-OCD connection, 20/No picocom output, 28/No noticeable change in all applications, 8/At least one application changed}}
\end{tikzpicture}
    \caption{Results of the 1024 bit flips.}
    \label{fig:pie_chart}
    \vspace{-0.5cm}
\end{figure}

In our analysis, we identified two significant leakages for the ChaCha encryption application caused by single bit flips in the LL key. ChaCha, which uses a 256-bit key to encrypt plaintext, experienced these vulnerabilities due to disruptions in the compressed decoder triggered by incorrect LL key bits. The first leakage resulted in 50\% of the encryption key being transmitted to the UART output, while the second leakage exposed the entire encryption key. Due to space constraints in this research article, we will only delve deeply into the second, more critical vulnerability.

When analyzing the netlist of the MiG-V to elaborate on the complete encryption key leakage over the ciphertext output at the UART output, we identified the key gate. The key gate is located in the compressed decoder and connected to bit 5 of the compressed instruction. Within the decoder tree, the key gate can modify the interpretation of the compressed instructions for logic operations. The relevant RISC-V compressed instructions for that bit flip are depicted in Table~\ref{tab:riscv_encoding}. 
\begin{table}[]
    \centering
    \begin{tabular}{c|c|c|c|c|c|c|c|c|c|>{\bfseries}c|c|c|c|c|c||c}
        \multicolumn{16}{c|}{Encoding bit} & \multirow{2}{*}{Instruction}\\
         15 & 14& 13 & 12 & 11& 10 & 9& 8& 7& 6& 5& 4& 3& 2& 1& 0 & \\\hline \hline
         1 & 0& 0 & 0 & 1 & 1 & \multicolumn{3}{c|}{rs1'/rd'} & 1 & 0 & \multicolumn{3}{c|}{rs2'}  & 0 & 1 & C.OR \\ \hline
         1 & 0& 0 & 0 & 1 & 1 & \multicolumn{3}{c|}{rs1'/rd'} & 1 & 1 & \multicolumn{3}{c|}{rs2'}  & 0 & 1 & C.AND \\ 

    \end{tabular}
    \caption{RISC-V bit encoding of the compr. instructions C.OR and C.AND, differing in a single bit.}
    \label{tab:riscv_encoding}
    \vspace{-0.3cm}
\end{table}
Flipping the bit by changing the LL key bit results in every C.OR operation being interpreted as a C.AND operation and vice versa. For the boot process of the processor, this does not result in any changes. However, for cryptographic algorithms, switching logic operations can have a significant impact on functionality and thus security. Specifically, for the ChaCha operation, it impacts the rotation function used for the encryption. The ChaCha algorithm works as follows:
\begin{enumerate}
    \item Determine an initial vector matrix (see Table~\ref{tab:IV}).
    \item The vector matrix is used to generate the key stream using mixing and rotation operations forming a single encryption round. An example of the rotation mechanism is depicted in Table~\ref{tab:rotation} for an input $p=0x12345678$.
    \item Once the 20 rounds are completed, the resulting keystream can be XORed with the plaintext.
\end{enumerate}
\begin{table}
    \centering
        \begin{tabular}{c|c|c|c}
            \hline
            \multicolumn{4}{c}{Initializing Vector Matrix} \\ 
            \hline
             nonce &  nonce &  nonce &  nonce \\ \hline
             key[0] &  key[1] &  key[2] &  key[3]  \\ \hline
             key[4] &  key[5] & key[6] &  key[7]  \\ \hline
             counter &  counter &  counter & counter \\ 
        \end{tabular}
        \caption{Structure of $IV^{0}$ from ChaCha}
        \label{tab:IV}
        \vspace{-0.4cm}
\end{table}

\begin{table}
    \centering
        \begin{tabular}{c|c|c|c|c}
            \hline
            \multicolumn{5}{c}{ChaCha mixing and rotation}\\ \hline
            \multirow{2}{*}{\begin{tabular}[c]{@{}c@{}}Rotation function of ChaCha \\ with input p = 0x12345678\end{tabular}} & a & b & \multirow{2}{*}{a \textbar{} b} & \multirow{2}{*}{a \& b}\\ 
            \cline{2-3}
                & p \(<\!\!<\) n & p \(>\!\!>\) (32-n) &  &\\ \hline
                n=16 & 0x56780000 & 0x00001234 & 0x56781234 & 0x00000000  \\ \hline
                n=12 & 0x45678000 & 0x00000123 & 0x45678123 & 0x00000000  \\ \hline
                n=8 & 0x34567800 & 0x00000012  & 0x34567812 & 0x00000000  \\ \hline
                n=7 & 0x1a2b3c00 & 0x00000009  & 0x1a2b3c09 & 0x00000000  \\ 
        \end{tabular}
        \caption{Rotation of ChaCha with intended OR and transformed (switching c.OR and c.AND)}
        \label{tab:rotation}
        \vspace{-0.4cm}
\end{table}

However, although the rotation and mixing are conducted as intended when using the C.OR operation, the mixing results in $0x00000000$ values when using a C.OR operation, as depicted in the right column of Table~\ref{tab:rotation}. This manipulation results in the mixed vector matrix $IV^{20}$ as illustrated in Table~\ref{tab:iv20}. The replacement of C.OR operations with C.AND operations leads to two rows filled with zeros, and a row storing half of the unmodified encryption key. When adding the initial vector matrix to the final matrix, the operations yield the values depicted in Table~\ref{tab:keystream}. 

\begin{table}
    \centering
        \begin{tabular}{c|c|c|c}
            \hline
            \multicolumn{4}{c}{Initializing Vector Matrix after 20 Mixing Rounds} \\ 
            \hline
            mixed nonce & mixed nonce & mixed nonce & mixed nonce \\ \hline
            0x00000000 & 0x00000000 & 0x00000000 & 0x00000000  \\ \hline
            key[4] & key[5] & key[6] & key[7]  \\ \hline
            0x00000000 & 0x00000000 & 0x00000000 & 0x00000000 \\ 
        \end{tabular}
        \caption{Structure of $IV^{20}$ from ChaCha with Trojan key (switching c.OR and c.AND)}
        \label{tab:iv20}
        \vspace{-0.5cm}
\end{table}

\begin{table}
    \centering
        \begin{tabular}{c|c|c|c}
            \hline
            \multicolumn{4}{c}{Keystream ($IV^{0}$ + $IV^{20}$)} \\ 
            \hline
            mixed nonce & mixed nonce & mixed nonce & mixed nonce \\ \hline
            key[0] & key[1] & key[2] & key[3]  \\ \hline
            key[4] * 2 & key[5] * 2 & key[6] * 2 & key[7] * 2  \\ \hline
            counter & counter & counter & counter \\ 
        \end{tabular}
        \caption{ChaCha resulting keystream with Trojan key (switching c.OR and c.AND)}
        \label{tab:keystream}
        \vspace{-0.5cm}
\end{table}

As the keystream is now only XORed to the plaintext, the encryption key is simply written to the output when using plaintexts that are entirely filled with zeros.

\subsection{Trojan Area Results}
Table~\ref{tab:table_usage_after} displays the resource utilization on the FPGA after the Trojan insertion. The data indicates that the increase in resource consumption is minimal. Most components, such as e.g. LUTRAMs and BRAMs are not significantly impacted by the Trojan implementation.
The only notable changes are in the number of LUTs (Look-Up Tables) and FFs (Flip-Flops), which increase by just 0.47\% and 0.21\%, respectively. This minor increase underscores such a Trojan's ability to conceal itself in a complex hardware design while having a major impact on the security of the system. 
To further reduce resource usage, the number of attack keys stored in the Trojan can be minimized. During the attack design phase, a large number of keys can be initially accommodated. Once the relevant assault vectors are identified, the key count can be trimmed to the essential quantity required for the attack.

\begin{table}[h!]
  \begin{center}
  \vspace{-0.4cm}
    \caption{Resource utilization of SoC after Trojan insertion.}
    \label{tab:table_usage_after}
    \begin{tabular}{l|c|c|c|r}
      \toprule 
      \textbf{Resource} & \textbf{Utilisation} & \textbf{Available} & \textbf{Utilisation \%} & \textbf{Change \%}\\
      \midrule 
      LUT & 46731 & 274080 & 17.05 & +0.47\\
      LUTRAM & 800 & 144000 & 0.56 & 0\\
      FF & 42736 & 548160 & 7.80 & +0.21\\
      BRAM & 62.5 & 912 & 6.85 & 0\\
      IO & 71 & 328 & 21.65 & 0\\
      MMCM & 2 & 4 & 50 & 0\\
      PLL & 1 & 8 & 12.5 & 0\\
      \bottomrule 
    \end{tabular}
  \end{center}
  \vspace{-1.2cm}
\end{table}

\section{Limitations}
\label{ch:discussion}

Our study reveals significant vulnerabilities in the MiG-V processor caused by LL, yet it has some limitations. We only analyzed single bit flips within a 1024-bit LL key, which does not capture the complex interactions from multiple concurrent bit flips. Additionally, our vulnerability detection was done manually, which is time-consuming and prone to false negatives, highlighting the need for automated tools to provide a more systematic and exhaustive evaluation. This suggests that additional vulnerabilities caused by LL may exist in the MiG-V. Despite these limitations, we highlight critical weaknesses in current LL schemes that could jeopardize data confidentiality, underlining the necessity for more comprehensive analyses of security properties post-LL.

\section{Possible Mitigation}
Information flow analysis tools allow tracking the flow of sensitive data when analyzing a hardware description~\cite{qflow, qflow2, qflow3}. The analysis can be integrated into the logic locking flow to determine whether the logic locking framework introduced any new hardware that allows unintended data leakages. If the framework identifies leakages, the logic locking process is interrupted and the process is restarted until the locking process is complete without introducing any vulnerabilities.
Moreover, the hardware designer can label sensitive signals, which are then avoided by the logic locking framework during the locking process.
\section{Conclusion}
Logic locking is designed to secure hardware by protecting integrated circuits from malicious alterations, known as hardware Trojans. For the first commercially available logic-locked RISC-V processor, MiG-V, we evaluated the impact of the logic locking scheme on the confidentiality of sensitive data in cryptographic software running on the processor. Our findings indicate that logic locking can significantly threaten the confidentiality of sensitive data in the MiG-V. This study is the first to identify these new security vulnerabilities caused by logic locking schemes.
We demonstrated that the logic locking scheme Inter-Lock introduces vulnerabilities that lead to data leakages in cryptographic algorithms running on the MiG-V under malicious logic locking keys. Specifically, by altering the logic locking to a malicious one, 100\% of the encryption key is forwarded to the hardware's output instead of the ciphertext.

In future research, this analysis could be expanded to systems running a secure kernel, such as seL4, to assess the security impact on complete TLS communications using encryption algorithms.
 
\ack{We would like to thank Hensoldt Cyber GmbH for providing us with the MiG-V Evaluation board and the necessary technical data sheets describing the software and hardware setup.}

\bibliography{bibtex.bib}

\end{document}